\documentclass[showpacs,12pt,prc,preprint]{revtex4-1}
\usepackage{indentfirst}
\usepackage[mathscr]{eucal}
\usepackage[dvips]{graphicx}
\usepackage{subeqnarray}
\def\lsim{\mathrel{\raise2pt\hbox to 8pt{\raise -5pt\hbox{$\sim$}\hss{$<$}}}}

\def\ds{\displaystyle}
\def\c{{\rm c}}
\def\ii{{\rm i}}

\begin{document}

\title{Spin and pseudospin symmetries in the antinucleon spectrum of nuclei}

\author{R. Lisboa}
\affiliation{Escola de Ci\^encias e Tecnologia, Universidade Federal do Rio Grande do Norte,
59014-615 Natal, Rio Grande do Norte, Brazil}
\affiliation{Departamento de F\'isica, Instituto Tecnol\'ogico de Aeron\'autica,
Centro T\'ecnico Aeroespacial, 12228-900 S\~ao Jos\'e dos Campos, S\~ao Paulo, Brazil}

\author{M. Malheiro}
\affiliation{Departamento de F\'isica, Instituto Tecnol\'ogico de Aeron\'autica,
Centro T\'ecnico Aeroespacial, 12228-900 S\~ao Jos\'e dos Campos, S\~ao Paulo, Brazil}

\noaffiliation

\author{P. Alberto}
\author{M. Fiolhais}
\affiliation{Departamento de F\'isica and Centro de F\'isica
Computacional, Universidade de Coimbra, P-3004-516 Coimbra,
Portugal}

\author{A. S. de Castro}
\affiliation{Departamento de F\'isica e Qu\'imica,
Universidade Estadual Paulista, 12516-410 Guaratinguet\'a, S\~ao
Paulo, Brazil}

\date{\today}

\begin{abstract}
{  Spin and pseudospin symmetries in the spectra of nucleons and antinucleons
are studied in a relativistic mean-field theory with scalar and vector Woods-Saxon potentials,
in which the strength of the latter is allowed to change.
We observe that, for nucleons and antinucleons, the spin symmetry is of perturbative nature and it is almost
an exact symmetry in the physical region for antinucleons. The opposite situation is found in
the pseudospin symmetry case, which is better realized for nucleons than for antinucleons, but is of dynamical nature and cannot
be viewed in a perturbative way both for nucleons and antinucleons.
This is shown by computing the spin-orbit and pseudospin-orbit couplings for selected spin and pseudospin partners in both spectra.
}
\end{abstract}

\maketitle

\section{Introduction}

Spin and pseudospin symmetries of the Dirac equation with scalar, $S$,
and vector, $V$, potentials are observed, respectively, when the difference, $\Delta=V-S$, or the sum, $\Sigma=V+S$, are constants.
These constants are zero for bound systems whose potentials go to zero at infinity.
Generally, in physical systems with this kind of potentials neither of these conditions is met exactly but, in some cases, one of them can be approximately true. As Ginocchio pointed out, these symmetries may explain degeneracies in some heavy meson spectra (spin symmetry) or in single-particle energy levels in nuclei
(pseudospin symmetry), when these physical systems are described by relativistic mean-field theories with scalar and vector potentials
\cite{gino_rev_2005}.
In terms of the non-relativistic quantum numbers $n\,l\,j$
(i.e., the quantum numbers of the upper component of the Dirac spinor),
exact spin symmetry means that the doublets $(n,l,j=l-1/2)$ $(n,l,j=l+1/2)$
are degenerate (no spin-orbit coupling) whilst in the case of pseudospin symmetry the degeneracy refers to
the doublets $(n',l+2,j=l-1/2)$ $(n,l,j=l+1/2)$.
In the latter case, a new principal quantum number, $\tilde n$, and a new orbital angular momentum
quantum number, $\tilde l$, are defined, such that the doublets are labeled by
$(\tilde{n},\tilde{l},\tilde{j}=\tilde{l}\pm 1/2)$. While $n,\,l$ are the common quantum numbers of
the upper component of spin symmetry doublets, $\tilde n,\,\tilde l$ are the common quantum numbers of
the lower component of the spinor of the  pseudospin symmetry doublets. The relations between $n'$, $\tilde{n}$ and $n$ depend on the shape
of the central mean-field. In the case of nuclei, with Woods-Saxon-like potentials, they are given by
$n'=n-1,\,\tilde n= n-1$.

The study of the antinucleon spectrum in nuclei is of interest because of the conjectures about the existence of
antinucleon bound levels inside the nucleus based on the relativistic description of nucleons \cite{burvenich_2002}.
Therefore, in this paper we are going to examine in some detail the onset of spin and pseudospin symmetries in
antinucleon single-particle states and compare it to what happens in the nucleon spectra.
It has been shown \cite{gin_ho,Zhou,prc_73_054309} that pseudospin and spin symmetry are connected by charge conjugation since, under this operation, the sign of the vector potential $V$ is changed
while the sign of the scalar potential $S$ remains the same, thus converting $\Delta$ into $-\Sigma$
and $\Sigma$ into $-\Delta$.
This was shown explicitly for harmonic oscillator potentials \cite{prc_73_054309}. Zhou \textit{et al.}
\cite{Zhou} have performed a realistic self-consistent mean-field calculation for antinucleons,  showing
how the shallow negative $\Sigma$ nucleon binding potential gives rise to a deep negative $-\Delta$ antinucleon binding potential, thereby holding many antinucleon single-particle states.

In this paper we investigate charge-conjugation effects in spin and pseudospin symmetries in nuclei,
in an unified way, by computing the energies of single-particle states of neutrons and antineutrons using scalar and vector Woods-Saxon potentials. The antineutron states are classified exactly in the same way as the neutron states, namely by the quantum numbers of the respective spinor upper component, and the spin and pseudospin partners obtained accordingly. In this study we keep constant
the range and diffusivity parameters that fit the neutron spectrum of $^{208}$Pb, used in previous works
\cite{prl_86_5015,prc_65_034307}, as well as the depth of the scalar potential, but vary the height of the vector potential. From this systematic study we can assess the nature of the spin and pseudospin symmetries for nucleons and antinucleons, namely its perturbative or non-perturbative nature. This is done by computing the contributions of the spin-orbit and pseudospin-orbit couplings to the energy
splittings of spin and pseudospin partners, both for neutrons and antineutrons. A comparison between the radial functions of
those partners is also made.

This paper is organized as follows. In Sec.~\ref{sec:dirac_cc_gen} we
present the general features of charge conjugated solutions of the Dirac equation with spherical scalar and vector
potentials and discuss their quantum numbers and their relation with pseudospin quantum numbers.
In the subsequent section we present the numerical solutions of the Dirac equation for both particles and
antiparticles in Woods-Saxon mean-field scalar and vector potentials with parameters that best fit
the single-particle states of $^{208}$Pb, but allowing for a variable strength of the
vector potential. Finally, in Section \ref{Sec:conclusions}, we draw the conclusions.

\section{Charge conjugation in the Dirac Hamiltonian with scalar and vector radial potentials}
\label{sec:dirac_cc_gen}
The Dirac Hamiltonian for a particle with mass $m$ under
the action of external scalar, $S$, and vector, $V$,
potentials reads ($\hbar=c=1$)
\begin{equation}\label{eq:ham_dirac}
H=\vec\alpha\cdot\vec p + \beta (m + S) + V\ ,
\end{equation}
where $\vec\alpha$ and $\beta$ are the Dirac matrices. The
time--independent Dirac equation for fermions, with energy $E$, is
\begin{equation}\label{eq:mov_dirac}
H \psi = E \psi \ .
\end{equation}
The charge-conjugation operator is given by $C=\ii\gamma^2\,K$ where $K$  is the
complex conjugation operator \cite{itzuber}. When
$C$ is applied to both sides of the Eq.~(\ref{eq:mov_dirac}), one obtains
the conjugate Dirac equation
\begin{equation}\label{eq:mov_dirac_cc}
H_\c\psi_\c=-E_\c \, \psi_\c\ ,
\end{equation}
where the conjugated spinor
is given by $\psi_{\rm c}=\ii\gamma^2\,\psi^*$ and the conjugate Hamiltonian is
\begin{equation}\label{eq:ham_dirac_cc}
H_{\rm c}=\vec\alpha\cdot\vec p + \beta (m + S) - V \,,
\end{equation}
and $E_c$ is the total energy related to the charge-conjugated spinor.
From Eqs.~(\ref{eq:ham_dirac}) and (\ref{eq:ham_dirac_cc}) one concludes
that charge-conjugation operation changes the sign of the vector potential
while keeping the sign of the scalar potential \cite{prc_73_054309}.
Therefore, the Hamiltonian is invariant under charge conjugation
when there is only a scalar potential and, in such a case, the antifermion energies
are the symmetric of the corresponding fermion energies.

For spherical symmetric systems, the fermion Dirac spinor can be written as
\begin{equation}\label{eq:psi}
\psi =
\left(
\begin{array}{r}
\ii\, G_{n\kappa}(r)\, \phi_{\kappa m_j}(\theta,\varphi)\\[0.1cm]
    F_{n\kappa}(r)\,\vec\sigma\cdot\hat{r}\,\,\phi_{\kappa m_j}(\theta,\varphi)
\end{array}
\right)
=
\left(
\begin{array}{r}
\ii\, G_{n\kappa}(r)\,\,\phi_{ \kappa m_j}(\theta,\varphi)\\[0.1cm]
  - F_{n\kappa}(r)\,\,\phi_{-\kappa m_j}(\theta,\varphi)
\end{array}
\right)
\end{equation}
where $\phi_{\kappa m_j}(\theta,\varphi)$ are the spinor spherical harmonics
and $G_{n\kappa}(r)$ and $F_{n\kappa}(r)$ are the radial
wave functions for the upper and lower components.
The $\kappa$ quantum number is related to the total angular
momentum, $j$, and orbital angular momentum, $l$, through
\begin{equation}\label{eq:def_kappa}
\kappa=\left\{
\begin{array}{cl}
- (l+1) &               \quad j =  l + {1\over 2} \\
   l    &               \quad j =  l - {1\over 2}\, .
\end{array}
\right.
\end{equation}
Hence, $\kappa$ contains the information about both angular quantum
numbers $l,j$ which can be obtained from
$l=|\kappa|+\frac12\big(\kappa/|\kappa|-1\big)$
and
$j=|\kappa|-1/2\ .$
The values of $\kappa$ for the upper and lower spinor spherical harmonics are such that,
if the upper spinor has orbital angular momentum $l$, the lower spinor
orbital angular momentum should be $\tilde l=l-\kappa/|\kappa|$.

To describe the antinucleons we obtain from Eq.~(\ref{eq:psi})
the corresponding conjugate spinor
\begin{equation}\label{eq:psi_cc}
\psi_\c=\ii\gamma^2\,\psi^*=
\left(
\begin{array}{r}
- F_{n\kappa} (r)\,\ii\sigma_2\, \, \phi_{-\kappa m_j}^*(\theta,\varphi)\\[0.1cm]
\ii\, G_{n\kappa} (r)\,\ii\sigma_2\, \, \phi_{\kappa m_j}^*(\theta,\varphi)
\end{array}
\right)\ ,
\end{equation}
which can still be written as
\begin{equation}\label{eq:psi_c_radial}
\psi_{\rm c} =
(-1)^{m_j-(\kappa/|\kappa|)/2+1}
\,\ii \, \,
\left(
\begin{array}{r}
\ii\,F_{n\kappa}(r)\,\phi_{-\kappa -m_j}(\theta,\varphi)\\[0.1cm]
  -G_{n\kappa}(r)\,\phi_{ \kappa -m_j}(\theta,\varphi)
\end{array}
\right)
\end{equation}
or, after performing the replacements
\begin{subeqnarray}
-\kappa&\rightarrow&\bar\kappa\\
F_{n\kappa} (r)&\rightarrow&\bar G_{\bar{n}\bar\kappa} (r)\\[0.1cm]
G_{n\kappa} (r)&\rightarrow&\bar F_{\bar{n}\bar\kappa} (r)\ ,
\end{subeqnarray}

\begin{equation}
\psi_\c=
(-1)^{m_j+(\bar{\kappa}/|\bar{\kappa}|)/2+1}
\,\, \ii \, \,
\left(
\begin{array}{r}
\ii\,\bar G_{\bar{n}\bar{\kappa}}(r)\,\phi_{ \bar{\kappa}, -m_j}(\theta,\varphi)\\[0.1cm]
  -\bar F_{\bar{n}\bar{\kappa}}(r)\,\phi_{-\bar{\kappa}, -m_j}(\theta,\varphi)\,
\end{array}
\right)
\label{eq:psi_c_radial2}
\end{equation}
We may label the spinor $\psi_\c$ using the quantum numbers
of its upper component, namely
$\bar{n}\, \bar{l}\, {\bar{j}}$. These are to be regarded as
the quantum numbers for the charge conjugated spinor which is the solution
of the charge conjugated Dirac equation~(\ref{eq:mov_dirac_cc}).

For fermions, the first-order differential radial equations are obtained
from Eq.~(\ref{eq:mov_dirac}) in the usual way, leading to
\begin{eqnarray}\label{eq:mov_radiais}
G'_{n\kappa} + \frac{1+\kappa}r\, G_{n\kappa} & = & (E+m-\Delta)\,F_{n\kappa}\nonumber\\
F'_{n\kappa} + \frac{1-\kappa}r\, F_{n\kappa} & = &-(E-m-\Sigma)\,G_{n\kappa} \ .
\end{eqnarray}
where $\Delta=V-S$ and $\Sigma=V+S$.

We use the same procedure to obtain the radial differential equations
for antifermions from the conjugate Dirac Hamiltonian~(\ref{eq:mov_dirac_cc})
and spinor~(\ref{eq:psi_c_radial2}). One obtains
\begin{eqnarray}\label{eq:mov_radiais_cc}
\bar G'_{\bar{n}\bar\kappa} + \frac{1+\bar\kappa}r\,\bar G_{\bar{n}\bar\kappa} & = &-(E_\c-m-\Sigma)\,\bar F_{\bar{n}\bar\kappa}\nonumber\\
\bar F'_{\bar{n}\bar\kappa} + \frac{1-\bar\kappa}r\,\bar F_{\bar{n}\bar\kappa} & = & (E_\c+m-\Delta)\,\bar G_{\bar{n}\bar\kappa} \ .
\end{eqnarray}

We observe that these equations for antifermions are very similar to the above
Eqs.~(\ref{eq:mov_radiais}) for fermions, the main effect of charge conjugation being
the transformation $\Delta\rightarrow-\Sigma$ and $\Sigma\rightarrow -\Delta$ as already mentioned in the Introduction.
Moreover, there is an other important modification: the binding
energy for the fermions is given by $\epsilon=E-m$ while
the binding energy for antifermions is given by $\epsilon_\c=-E_\c-m$.
In terms of these eigenvalues, Eqs.~(\ref{eq:mov_radiais}) and (\ref{eq:mov_radiais_cc}) become, respectively,
\begin{eqnarray}\label{eq:mov_radiais_m}
G'_{n\kappa} + \frac{1+\kappa}r\,G_{n\kappa} & = & (\epsilon+2m-\Delta)\,F_{n\kappa} \nonumber\\
F'_{n\kappa} + \frac{1-\kappa}r\,F_{n\kappa} & = &-(\epsilon-\Sigma)\, G_{n\kappa} \
\end{eqnarray}
and
\begin{eqnarray}\label{eq:mov_radiais_ccm}
\bar G'_{\bar{n}\bar\kappa}+\frac{1+\bar\kappa}r\,\bar G_{\bar{n}\bar\kappa}&=&(\epsilon_\c+2m+\Sigma)\,\bar F_{\bar{n}\bar\kappa}\nonumber\\
\bar F'_{\bar{n}\bar\kappa}+\frac{1-\bar\kappa}r\,\bar F_{\bar{n}\bar\kappa}&=&-(\epsilon_\c+\Delta)\,\bar G_{\bar{n}\bar\kappa} \ .
\end{eqnarray}
It is also instructive to write the second-order differential equations for both the upper and lower components of the fermions and antifermions spinors:
{\arraycolsep 1pt
\begin{eqnarray}\label{eq:2_ordem_mov_radiais_m_G}
G\,''_{n\kappa} &-&\frac{\kappa (\kappa +1)%
}{r^{2}}\,G_{n\kappa}+\frac{\Delta ^{\prime }}{\epsilon+2m-\Delta}\biggl(%
\frac{{\rm d}\hfil}{{\rm d}r}+\frac{1+\kappa }{r}\biggr)G_{n\kappa}
 =-(\epsilon-\Sigma)(\epsilon+2m-\Delta)G_{n\kappa}\\
 \label{eq:2_ordem_mov_radiais_m_F}
F\,''_{n\kappa} &-&\frac{\kappa (\kappa -1)%
}{r^{2}}\,F_{n\kappa}+\frac{\Sigma ^{\prime }}{\epsilon-\Sigma}\biggl(%
\frac{{\rm d}\hfil}{{\rm d}r}+\frac{1-\kappa }{r}\biggr)F_{n\kappa}
 =-(\epsilon-\Sigma)(\epsilon+2m-\Delta)F_{n\kappa} \,,
\end{eqnarray}
}
and
{\arraycolsep 1pt
\begin{eqnarray}\label{eq:2_ordem_mov_radiais_ccm_G}
\bar G\,''_{\bar{n}\bar\kappa}&-&\frac{\bar\kappa (\bar\kappa +1)%
}{r^{2}}\,\bar G_{\bar{n}\bar\kappa}-\frac{\Sigma^{\prime }}{\epsilon_\c+2m+\Sigma}\biggl(%
\frac{{\rm d}\hfil}{{\rm d}r}+\frac{1+\bar\kappa }{r}\biggr)\bar G_{\bar{n}\bar\kappa}
 =\nonumber\\
&&\hspace{7.5cm}=-(\epsilon_\c+\Delta)(\epsilon_\c+2m+\Sigma)\,\bar G_{\bar{n}\bar\kappa}\\
\label{eq:2_ordem_mov_radiais_ccm_F}
\bar F\,''_{\bar{n}\bar\kappa}&-&\frac{\bar\kappa (\bar\kappa -1)%
}{r^{2}}\,\bar F_{\bar{n}\bar\kappa}-\frac{\Delta ^{\prime }}{\epsilon_\c+\Delta}\biggl(%
\frac{{\rm d}\hfil}{{\rm d}r}+\frac{1-\bar\kappa }{r}\biggr)\bar F_{\bar{n}\bar\kappa}
 =-(\epsilon_\c+\Delta)(\epsilon_\c+2m+\Sigma)\bar F_{\bar{n}\bar\kappa} \,.
\end{eqnarray}
}
From these sets of equations it is even clearer the role interchange
of $\Delta$ and $\Sigma$ in the equations for the upper and
lower radial wave functions for fermions and antifermions. For instance, the
role of the $\Delta$ potential, which contributes to the effective mass
$m^*$ of fermions ($2m^*=\epsilon+2m-\Delta$), is  played by $-\Sigma$
in the case of antifermions, with a  ``conjugate'' effective mass such that $2m_\c^*=\epsilon_\c+2m+\Sigma$.
On the other hand, the role of the $\Sigma$ potential as the binding potential for fermions is now played by
$-\Delta$ for antifermions. We will discuss in the next section the consequences of this role change on the
level distribution and on the onset of the symmetries of the antinucleon spectrum.

Actually, the sets of equations (\ref{eq:mov_radiais_m}) and
(\ref{eq:mov_radiais_ccm}) are identical provided we make the correspondences $\Sigma\rightarrow -\Delta$, $\Delta\rightarrow -\Sigma$ and $\epsilon \rightarrow \epsilon_\c$ (corresponding to $E\rightarrow -E_\c$). Therefore, for a given pair $(S,V)$ of potentials the numerical results
for a certain positive energy state (the binding energy and wave functions) are the same as for a negative energy state
with the same quantum numbers and the pair $(S,-V)$, except, of course, for the total energy $E$ and $E_\c$.
We recall that the classification of the single-particle levels is given by the
quantum numbers of the upper component in both cases, hence a correspondence between the spin and
pseudospin partners with positive and negative energies can be made.

The second-order equations (\ref{eq:2_ordem_mov_radiais_m_G})-(\ref{eq:2_ordem_mov_radiais_ccm_F})
allow us to identify the spin- pseudospin-orbit terms for both fermions and antifermions
\cite{prc_65_034307}. These are
\[
-\frac{\Delta ^{\prime }}{\epsilon+2m-\Delta}%
\frac{1+\kappa }{r}\,G_{n\kappa}\quad {\rm (spin)}\qquad
-\frac{\Sigma ^{\prime }}{\epsilon-\Sigma}%
\frac{1-\kappa }{r}\,F_{n\kappa}\quad {\rm (pseudospin)}
\]
for fermions, and
\[
\frac{\Sigma^{\prime }}{\epsilon_\c+2m+\Sigma}%
\frac{1+\bar\kappa }{r}\,\bar G_{\bar{n}\bar\kappa}\quad {\rm (spin)}\qquad
\frac{\Delta ^{\prime }}{\epsilon_\c+\Delta}%
\frac{1-\bar\kappa }{r}\,\bar F_{\bar{n}\bar\kappa}\quad {\rm (pseudospin)}
\]
for antifermions. From these last expressions it is clear the role that $\Delta$ and $\Sigma$ potentials play, respectively, in the onset of the spin and pseudospin symmetries for fermions and the reversal of their roles for antifermions.
If one divides the second-order equations (\ref{eq:2_ordem_mov_radiais_m_G})-(\ref{eq:2_ordem_mov_radiais_ccm_F})
by (twice) the effective masses referred to above, one gets Schroedinger-like equations (see ref.~\cite{prc_65_034307}) thereby providing a way of obtaining the contributions of their various terms to the single-particle binding energy, $\epsilon$. In particular, one gets for the spin- and pseudospin-orbit contributions for the energy of the level with quantum numbers $n\kappa$ (fermions)
\begin{eqnarray}
\label{E_spin_pspin-orbit}
  E^{\rm SO}_{n\kappa}&=& - \frac{\ds\int_0^\infty\frac{\Delta ^{\prime }}{(\epsilon_{n\kappa}+2m-\Delta)^2}\frac{1+\kappa }{r}\,
  |G_{n\kappa}|^2\,r^2\,dr}{\rule{0pt}{6mm}\ds\int_0^\infty  |G_{n\kappa}|^2\,r^2\,{\rm d}r} \nonumber\\
  E^{\rm PSO}_{n\kappa}&=& - \frac{\ds\int_0^\infty\frac{\Sigma ^{\prime }}
  {(\epsilon_{n\kappa}-\Sigma)(\epsilon_{n\kappa}+2m-\Delta)}%
\frac{1-\kappa }{r}\,
  |F_{n\kappa}|^2\,r^2\,dr}{\rule{0pt}{6mm}\ds\int_0^\infty  |F_{n\kappa}|^2\,r^2\,{\rm d}r}\ .
\end{eqnarray}
In these formulas, the integration is taken in the principal value sense whenever the
denominators of the integrands are zero. This is the case for
$E^{\rm PSO}$ because $\epsilon-\Sigma$ is zero for some value of $r$. For antifermions, the formulas are
similar:
\begin{eqnarray}
\label{E_spin_pspin-orbit_cc}
  E^{\rm SO}_{\bar{n}\bar\kappa}&=& \frac{\ds\int_0^\infty\frac{\Sigma^{\prime }}{(\epsilon_{\bar{n}\bar\kappa}+2m+\Sigma)^2}\frac{1+\bar\kappa }{r}\,
  |\bar G_{\bar{n}\bar\kappa}|^2\,r^2\,{\rm d}r}
  {\rule{0pt}{6mm}\ds\int_0^\infty  |\bar G_{\bar{n}\bar\kappa}|^2\,r^2\,dr} \nonumber\\
  E^{\rm PSO}_{\bar{n}\bar\kappa}&=& \frac{\ds\int_0^\infty\frac{\Delta^{\prime }}
  {(\epsilon_{\bar{n}\bar\kappa}+\Delta)(\epsilon_{\bar{n}\bar\kappa}+2m+\Sigma)}%
\frac{1-\bar\kappa }{r}\,
  |\bar F_{\bar{n}\bar\kappa}|^2\,r^2\,dr}{\rule{0pt}{6mm}\ds\int_0^\infty  |\bar F_{\bar{n}\bar\kappa}|^2\,r^2\,{\rm d}r}\ .
\end{eqnarray}

We will use these formulas later on to assess the perturbative nature of the spin and pseudospin
symmetries for fermions and antifermions.

\section{Numerical results and discussion}

In previous works we used realistic mean-field Woods-Saxon potentials in the Hamiltonian~(\ref{eq:ham_dirac})
to study the structure of the neutron
single-particle spectrum of $^{208}$Pb \cite{prl_86_5015,prc_65_034307}.
By varying the parameters of the Woods--Saxon potentials, namely its depth, diffusivity and
range, we were able to perform a systematic investigation of the pseudospin energy splittings as a function
of those parameters. We concluded that the onset of the pseudospin symmetry in nuclei is dynamical,
since it results mainly from cancelations of several terms contributing to the single-particle levels,
instead of being a consequence of a too small pseudospin-orbit coupling. An equivalent statement is that the pseudospin
symmetry in nuclei is non-perturbative, as it was pointed out by other authors \cite{marcos}.

In the present work we follow a similar strategy, using mean-field Woods--Saxon potentials
whose diffusivity and range are adjusted to reproduce the neutron single-particle spectrum of $^{208}$Pb,
but allowing for the vector potential strength $V_0$ to vary, and study the resulting spectra for both neutrons and antineutrons.
We carry out this program by solving numerically the first-order Dirac equations for fermions and antifermions, i.e. Eqs.
(\ref{eq:mov_radiais_m}) and (\ref{eq:mov_radiais_ccm}), and obtaining the radial
wave functions and eigenenergies.
%

We used the following parameters, adjusted to $^{208}$Pb:  $S_{0}=-358.0$~MeV
for the depth of $S$, and $a_v=a_s=0.6$~fm and $R_v=R_s=7.0$~fm for, respectively, the diffusivity and range
of both $V$ and $S$ potentials \cite{prl_86_5015,prc_65_034307}. We vary $V_0$, covering a broad range
of values for the binding potential. The parameter $V_{0}=292.0$~MeV corresponds to the best fit to the neutron spectrum of $^{208}$Pb.

\begin{table}[!ht]
\caption{Single-particle neutron binding energies, in MeV, of two spin and two pseudo-spin partners
 for various strengths of the vector potential
 $V_0$. The column for $V_0= 292.0$ MeV
 corresponds to the best fit to the neutron single-particle energies of $^{208}$Pb.
 The states with zero binding energy are in the continuum.
         }
{\footnotesize
\begin{center}
  \begin{tabular*}{0.95\textwidth}%
     {@{\extracolsep{\fill}}lrrr|r|r}
  \hline
  \hline
$V_0$        &   0.0    &  100.0   &  200.0   \ &  292.0  \ &  300.0  \\
$\Sigma_0$   &-358.0    & -258.0   & -158.0   \ &  -66.0  \ &  -58.0  \\
$\Delta_0$   & 358.0    &  458.0   &  558.0   \ &  650.0  \ &  658.0  \\
\hline\hline
$\epsilon(2s_{1/2})$   &-322.5962 &-224.9267 &-128.2021 \ &-41.6087	\ &-34.3833 \\
$\epsilon(1d_{3/2})$   &-326.7516 &-228.7713 &-131.5698 \ &-44.0704	\ &-36.7021 \\
$\Delta E$   &   4.1554 &   3.8446 &   3.3677 \ &  2.4616	\ &  2.3188 \\
\hline
$\epsilon(2g_{9/2})$   &-255.2061 &-162.2730 &-72.7440  \ &-1.3198   \ &0.0   \\
$\epsilon(1i_{11/2})$  &-266.6147 &-172.3074 &-80.4942  \ &-3.3984   \ &0.0    \\
$\Delta E$   &  11.4086 &  10.0344 &  7.7502  \ & 2.0786   \ &$-$   \\
\hline
$\epsilon(1p_{1/2})$   &-338.3051 &-239.6125 &-141.4138 \ &-52.2731 	\ &-44.6683  \\
$\epsilon(1p_{3/2})$   &-338.4124 &-239.7783 &-141.6810 \ &-52.7714 	\ &-44.2065  \\
$\Delta E$   &   0.1073 &   0.1658 &   0.2672 \ &  0.4983   \ &  0.4618  \\
\hline
$\epsilon(1h_{9/2})$   &-283.3606 &-188.0162 &-94.6341  \ &-14.1499  \ & -7.9630   \\
$\epsilon(1h_{11/2})$  &-284.4187 &-189.6579 &-97.2570  \ &-18.5705  \ &-12.5662   \\
$\Delta E$   &   1.0581 &   1.6417 &  2.6229  \ &  4.4206  \ &  4.6031   \\
\hline
\hline
\end{tabular*}
 \end{center}
 \label{tab:nbelevel}
}
\end{table}


In Table~\ref{tab:nbelevel} the binding energies, $\epsilon$, and
the energy splittings, $\Delta E$, for two pseudospin and two spin partners
of  nucleon single-particle states  are shown, for a vector potential strength varying from
$V_0=0$ to $V_0=300.0$~MeV. As the
magnitude of $V_0$ increases, the magnitude (absolute value) of $\Sigma_0 = S_0 + V_0$
decreases and the energy splittings between pseudospin partners become smaller.
Concomitantly, the magnitude of $\Delta_0 =  V_0-S_0 $ increases and the
energies of the spin partners move away from each other. These results
are in agreement with those of previous works \cite{prc_65_034307} and they occur because, for nucleon systems,
the $\Sigma$ ($\Delta$)
potential drives the pseudospin- (spin)-orbit interaction. Moreover, one sees that,
since $\Sigma$ is the binding potential, for \mbox{$|\Sigma_0| < 66.0$~MeV},
the potential becomes too shallow and the higher energy states become unbound.

In Table \ref{tab:abelevel} we present the same quantities for antineutrons,
namely the binding energies, $\epsilon_\c$, and
the energy splittings for the same two spin and pseudospin
partners, using the same values of $V_0$ as in Table \ref{tab:nbelevel}.
This time, as the magnitude of $\Sigma_0$
decreases (and thus $\Delta_0$ increases) the energies of the spin partners become
quasi--degenerate whereas the energies of the pseudospin partners move away from each other
as was already remarked by Zhou \textit{et al.} \cite{Zhou}.
This can be readily explained by the reversal of the role played by the
$\Sigma$ and $\Delta$ potentials in the spin and pseudospin symmetries as explained before.
Contrary to the neutrons, for antineutrons
the energies of conjugate spin partners get almost degenerate
because of the smallness of the $\Sigma$ potential as compared to the $\Delta$ potential,
which drives the spin splitting for neutrons.

\begin{table}[!ht]
\caption{ The antineutrons binding energies, $\epsilon_\c$ (in MeV), and splittings for
two spin partners and two pseudospin partners.
The parameters are the same as in Table~\ref{tab:nbelevel}.
         }
{\footnotesize
\begin{center}
  \begin{tabular*}{0.95\textwidth}%
     {@{\extracolsep{\fill}}lrrr|r|r}
\hline   \hline
$V_0$                  &   0.0    & 100.0    & 200.0  \   & 292.0 \   & 300.0        \\
$\Sigma_0$             &-358.0    &-258.0    &-158.0  \   & -66.0 \   & -58.0       \\
$\Delta_0$             & 358.0    & 458.0    & 558.0  \   & 650.0 \   & 658.0        \\
\hline\hline
$\epsilon_\c(\bar{2}\bar{s}_{1/2})$  &-322.5962 &-420.7518 &-519.2080 \  &-609.9713 \ &-617.8704  \\ 
$\epsilon_\c(\bar{1}\bar{d}_{3/2})$  &-326.7516 &-425.1387 &-523.7798 \  &-614.6861 \ &-622.5967  \\ 
$\Delta E$              &   4.1554 &   4.3869 &   4.5718 \  &   4.7148 \ &   4.7263  \\
\hline
$\epsilon_\c(\bar{2}\bar{g}_{9/2})$  &-255.2061 &-349.7786 &-445.3284 \  &-533.8162 \ &-541.5317  \\  
$\epsilon_\c(\bar{1}\bar{i}_{11/2})$ &-266.6147 &-362.1573 &-458.4515 \  &-547.4968 \ &-555.2563  \\  
$\Delta E$              &  11.4086 &  12.3787 &  13.1231 \  &  13.6806 \ &  13.7246  \\
\hline
$\epsilon_\c(\bar{1}\bar{p}_{1/2})$  &-338.3051 &-437.2571 &-536.3716 \  &-627.6573 \ &-635.5988  \\ 
$\epsilon_\c(\bar{1}\bar{p}_{3/2})$  &-338.4124 &-437.3241 &-536.4082 \  &-627.6713 \ &-635.6111  \\ 
$\Delta E$              &   0.1073 &   0.0670 &   0.0366 \  &   0.0140 \ &   0.0123     \\
\hline
$\epsilon_\c(\bar{1}\bar{h}_{9/2})$  &-283.3606 &-380.3511 &-476.6347 \  &-566.1890 \ &-573.9897  \\  
$\epsilon_\c(\bar{1}\bar{h}_{11/2})$ &-284.4187 &-379.6942 &-476.9910 \  &-566.3244 \ &-574.1078  \\  
$\Delta E$              &   1.0581 &   0.6569 &   0.3563 \  &   0.1354 \ &   0.1181 \\
\hline
\hline
\end{tabular*}
 \end{center}
 \label{tab:abelevel}
}
\end{table}

Because of the correspondence that can be established between the radial equations for particles and for antiparticles
mentioned at the end of Sec. II., it is interesting to note that if we compute the energies and splittings for $V_0<0$, in the case of the neutrons, for a spin or pseudospin partner, we would get the same values as for the corresponding antineutron spin or pseudospin partner with the positive symmetric value $|V_0|$ and vice-versa (results for antineutrons with $V_0<0$ equal to the results for fermions with the symmetric value $|V_0|$). In other words, for the corresponding single-particle states of positive and negative energies the behaviour of neutrons for $V_0>0$ is the same {\em mutatis mutandis} as for antineutrons with $V_0<0$, and vice-versa. Actually, this correspondence can already be seen from the Tables \ref{tab:nbelevel} and \ref{tab:abelevel}, since for $V_0=0$ the single-particle states of neutrons and antineutrons with the same quantum numbers should have the same binding energies, and indeed this is what comes out from the calculations.

\begin{figure}[!ht]
\begin{center}
\includegraphics[width=10.5cm]{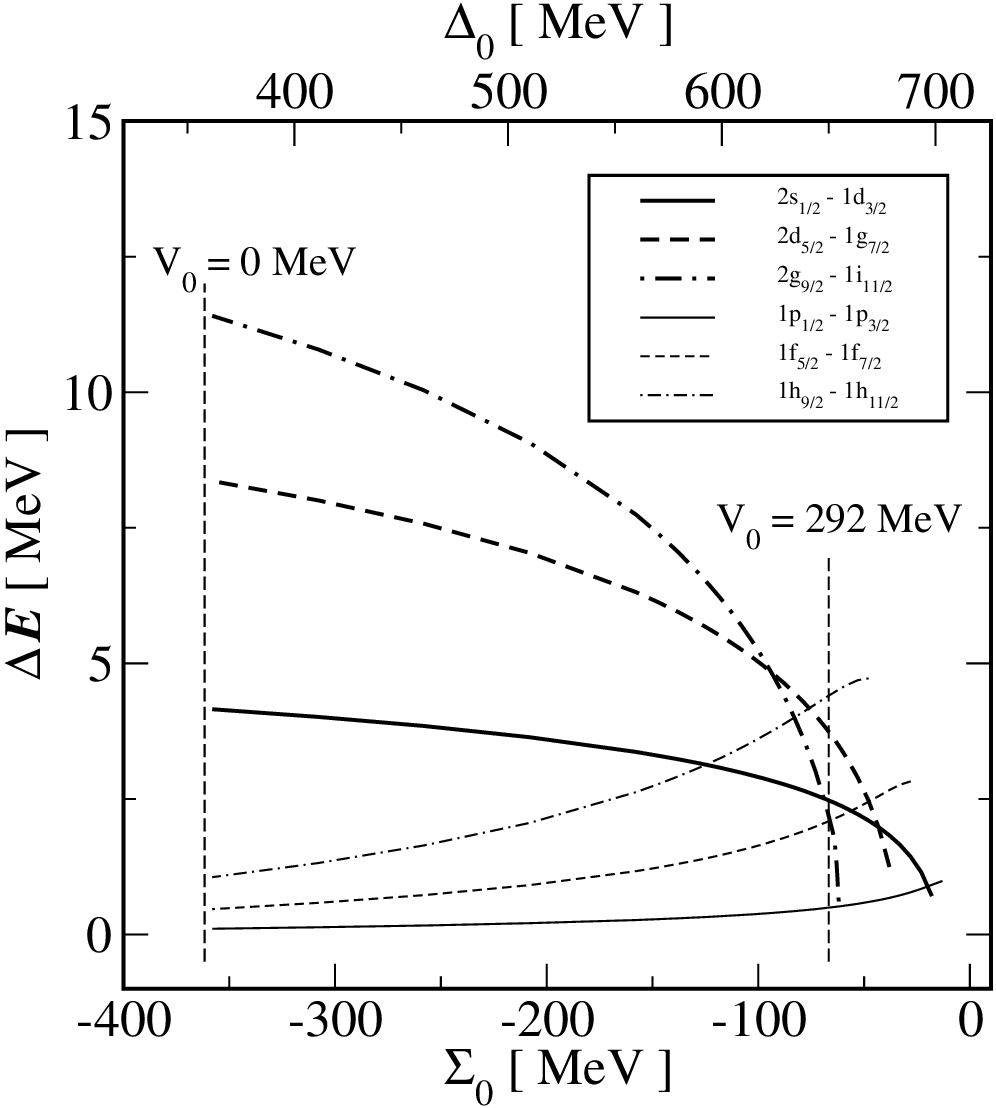}
\end{center}
\caption
{Splittings for spin and pseudospin partners for neutrons. The dashed line with $V_0=292$ MeV represents the physical region.}
\label{fig:deltaEneutrons}
\end{figure}

We now examine in a bit more detail both symmetries and their dependence on the central potentials $\Sigma$ and $\Delta$ for neutrons and
antineutrons. In Fig.~\ref{fig:deltaEneutrons} we show the splittings $\Delta E$ for three spin and pseudospin partners as a function of $\Sigma_0$, the binding potential depth. The vertical dashed line for $V_0=292$ MeV
stands for the parameters that best reproduce the experimental single-particle energies of $^{208}$Pb. For large absolute values of $|\Sigma_0|$ the spin symmetry is better realized because the strength of the $\Delta$ potential, $\Delta_0$, becomes smaller and thus the spin-orbit interaction gets weaker. As $|\Sigma_0|$ decreases, the spin-orbit interaction gets more and more important and the deviation from the exact symmetry gets bigger.
In the same figure we also represent the splittings for the pseudospin partners. The pseudospin symmetry never becomes an exact one, though the
splitting for the partners chosen stay close to zero in the physical region. But this is definitely not the case for large $\Sigma_0$. From the same figure one also concludes that the quality of both spin and pseudospin symmetries, as measured by the splittings, is similar for $^{208}$Pb.
However, there is clearly a different behaviour of the spin and pseudospin splittings, respectively, when $\Delta_0$ decreases and when $|\Sigma_0|$ decreases.

\begin{figure}[!ht]
{
\includegraphics[width=10.5cm]{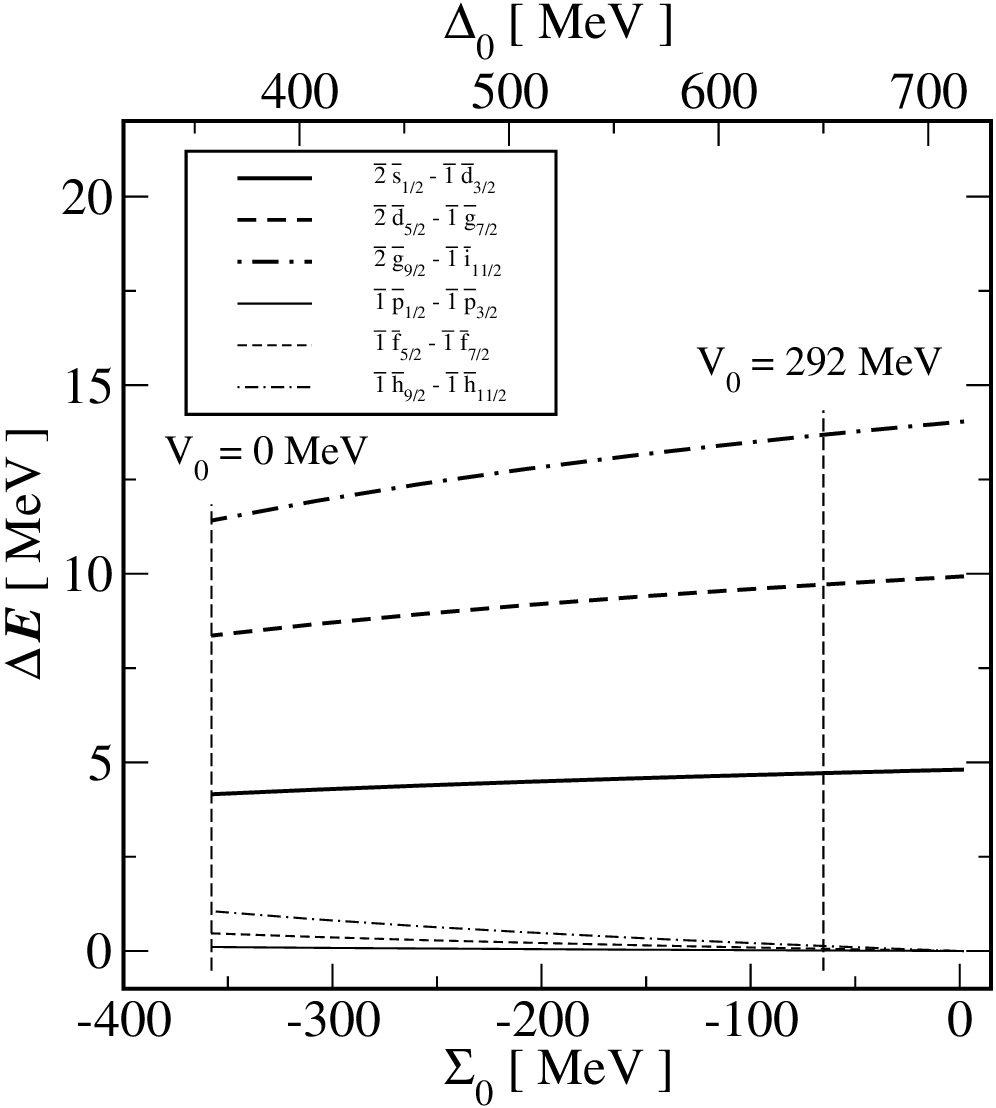}
\vspace*{-0.4cm}
}\hfill
\caption
{ Splittings for spin and pseudospin partners for antineutrons. The dashed line with $V_0=292$ MeV represents the physical region.
}
\label{fig:deltaEantineutrons}
\end{figure}

 A similar analysis for antineutrons can be made by looking at Fig.~\ref{fig:deltaEantineutrons}, where the splittings for three pseudospin and spin partners with the same quantum numbers as in Fig.~\ref{fig:deltaEneutrons} are shown again as a function of $\Sigma_0$.
One sees that for antineutrons the spin symmetry for small $\Sigma_0$ (physical region) is very good. On the contrary, pseudospin symmetry is
broken significantly. This, of course, was to be expected because of the roles that the $\Sigma$ and $\Delta$ potentials now play on the onset of
spin and pseudospin symmetries. From this figure and Fig.~\ref{fig:deltaEneutrons} one can also see that, as remarked before, for $V_0=0$ the
antinucleon levels
have the same splittings as the corresponding neutron levels. This is because is this case $\Sigma=-\Delta$ and therefore the charge-conjugation operation $\Delta\to -\Sigma\quad \Sigma\to -\Delta$ does not change the potentials. The total energy spectrum of the neutrons is then exactly symmetric to the total energy spectrum of the neutrons ($E_\c=-E)$.

\begin{figure}[!ht]
{
\includegraphics[width=10.5cm]{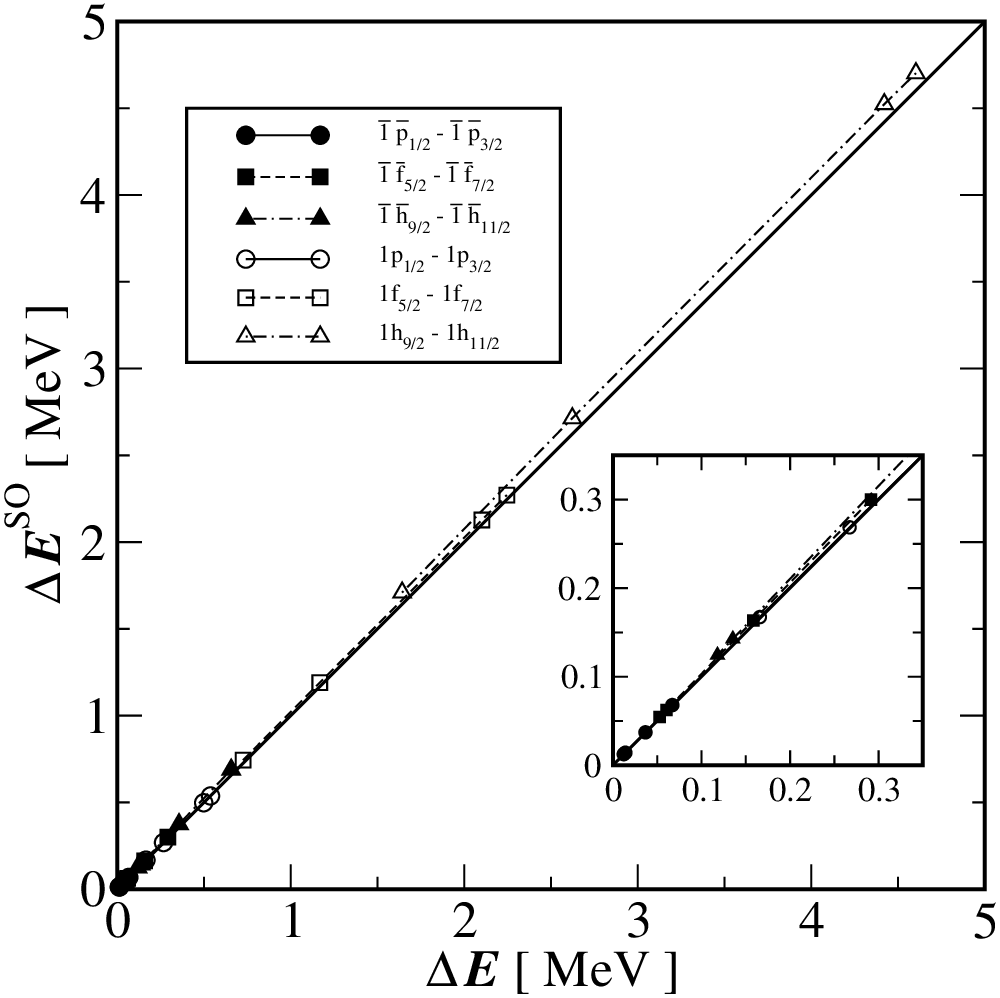}
\vspace*{-0.4cm}
}\hfill
\caption
{Spin-orbit terms splittings for three neutrons and antineutrons spin partners plotted against the respective energy splittings $\Delta E$.
The thicker solid line represents the values for which $\Delta E^{\rm SO}=\Delta E$.
}
\label{fig:spin_LS_DE}
\end{figure}

\begin{figure}[!ht]
{
\includegraphics[width=10.5cm]{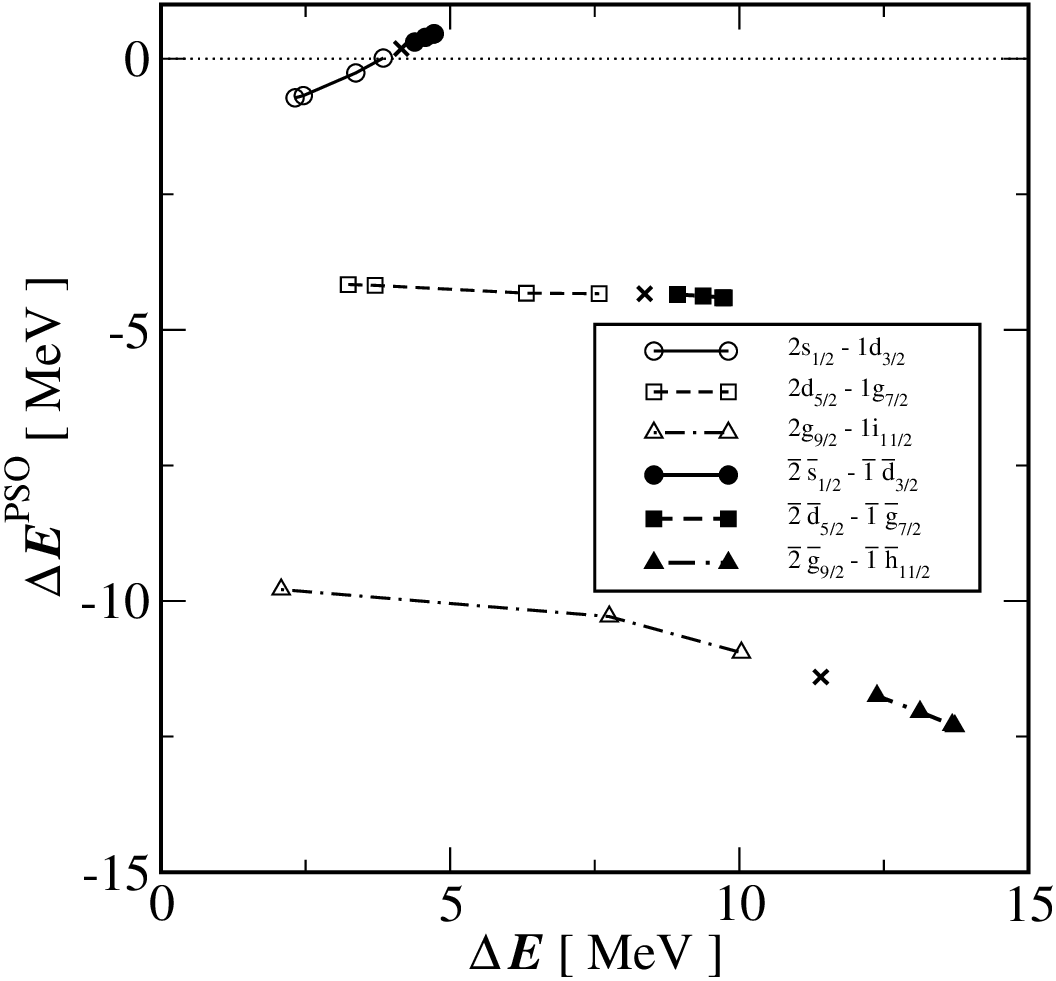}
\vspace*{-0.4cm}
}\hfill
\caption
{Pseudospin-orbit terms splittings for three neutrons and antineutrons spin partners plotted against
the respective energy splittings $\Delta E$. The points labeled by `\textsf{x}' mark the $V_0=0$ point for each pair of levels.}
\label{fig:pspin_LS_DE}
\end{figure}


In Figs.~\ref{fig:spin_LS_DE} and \ref{fig:pspin_LS_DE} is shown the splittings (i.e. the differences) of (P)SO (pseudo)spin-orbit terms
(\ref{E_spin_pspin-orbit}) and (\ref{E_spin_pspin-orbit_cc}) for three (pseudo)spin
partners, both for neutrons and antineutrons, for the four positive values of $V_0$ shown in tables \ref{tab:nbelevel} and \ref{tab:abelevel}.
From these figures one can clearly see the different nature of spin and pseudospin in nuclei, both for neutrons and antineutrons.
There is a correlation between the values the spin-orbit coupling and the energy splittings for spin partners,
the ratio $\Delta E^{\rm PS}/\Delta E$ being very close to 1 for antineutrons. This an unmistakeable sign of the perturbative nature of spin symmetry in nuclei, both for neutrons and antineutrons.

The situation for the pseudospin partners is completely different. There is no correlation between the pseudo-spin term splittings and the
energy splittings, even for small values of $\Delta E$. We see that even the sign is different in most cases.
Therefore, in spite of the fact, as was mentioned earlier,
that there is a connection between the strength of the $\Sigma$ ($\Delta$) potential and neutron (antineutron) pseudospin energy splittings,
there is not a relation between the respective pseudospin-orbit terms and the energy splittings, i.e., the onset of pseudospin symmetry.
Thus we can conclude that the pseudospin symmetry in nuclei is not perturbative, both for neutrons and antineutrons.

Finally, we present the radial functions $G$ and $F$ of one spin doublet and one pseudospin doublet.
We chose the pairs $[1f_{5/2}-1f_{7/2}]$ and $[2g_{7/2}-1i_{11/2}]$ because they have about the same energy
splittings for neutron levels for the physical parameters.

\begin{figure}[!ht]
{
\parbox[!ht]{8.2cm}{
\begin{center}
\includegraphics[width=8cm]{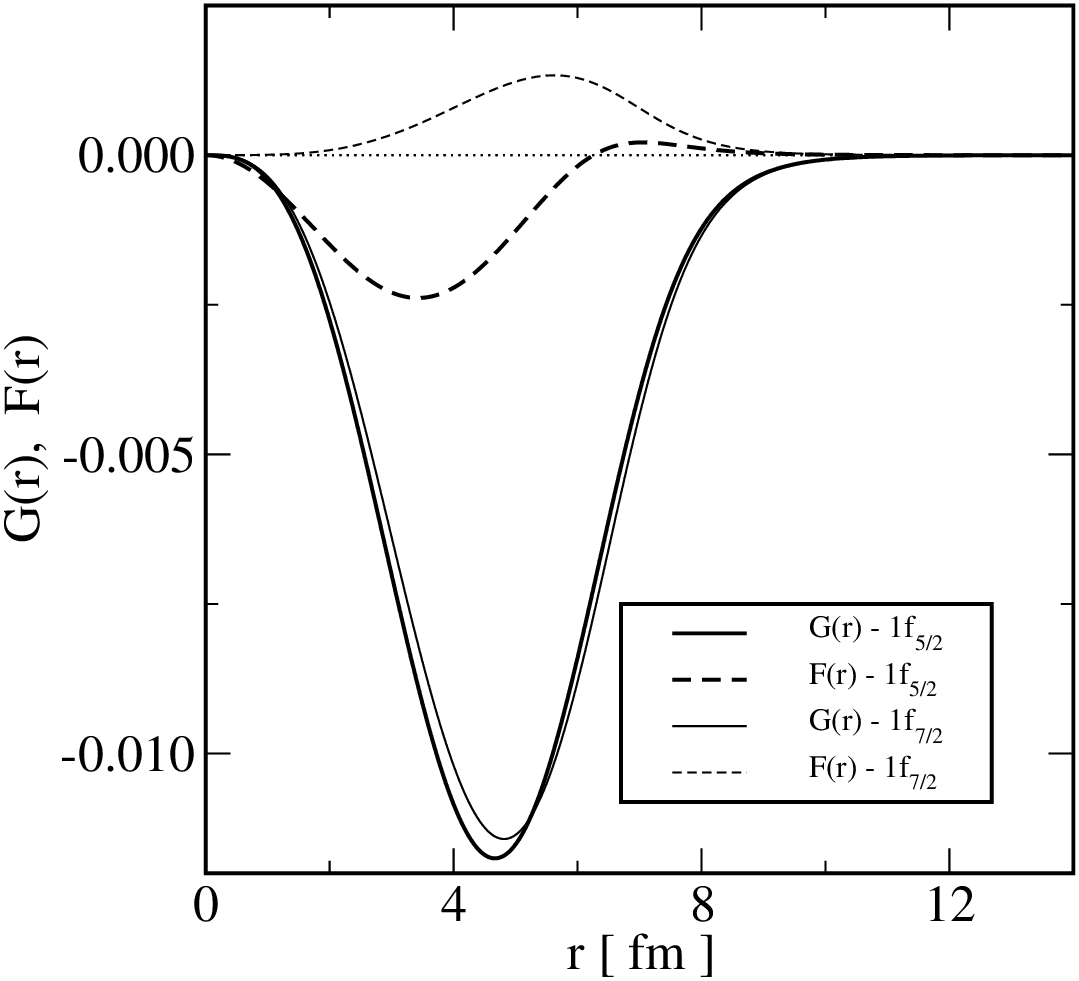}
\end{center}
}\hfill
\parbox[!ht]{8.2cm}{
\begin{center}
\includegraphics[width=8cm]{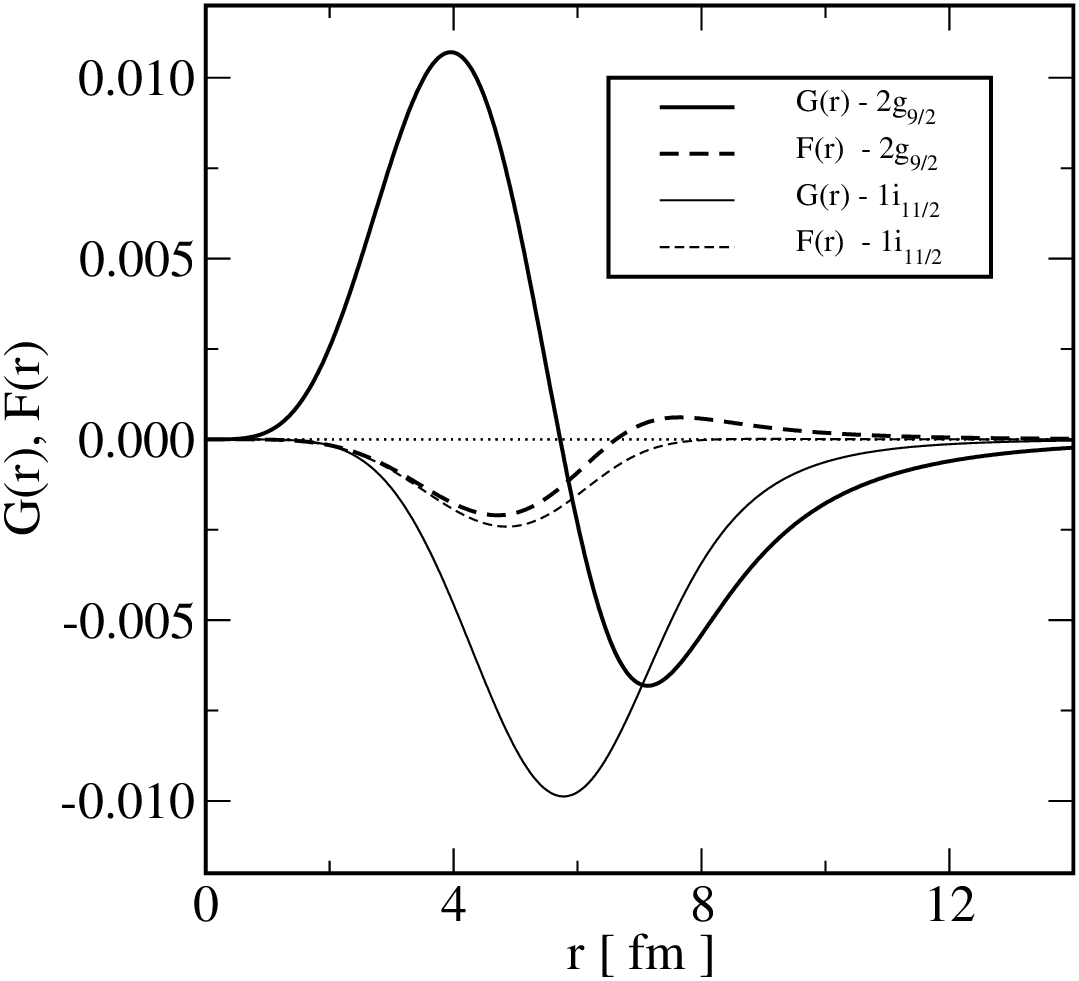}
\end{center} }  } \vspace*{-.2cm}
 \caption{Neutron radial wave functions $G$ and $F$ of the spin pair
 $1f_{5/2}-1f_{7/2}$ and the pseudospin pair
 $2g_{7/2}-1i_{11/2}$ for the physical
 parameters in scaled units.\hfill\ } \label{fig:F_G_spin_pspin_nn}
\end{figure}


\begin{figure}[!ht]
{
\parbox[!ht]{8.2cm}{
\begin{center}
\includegraphics[width=8cm]{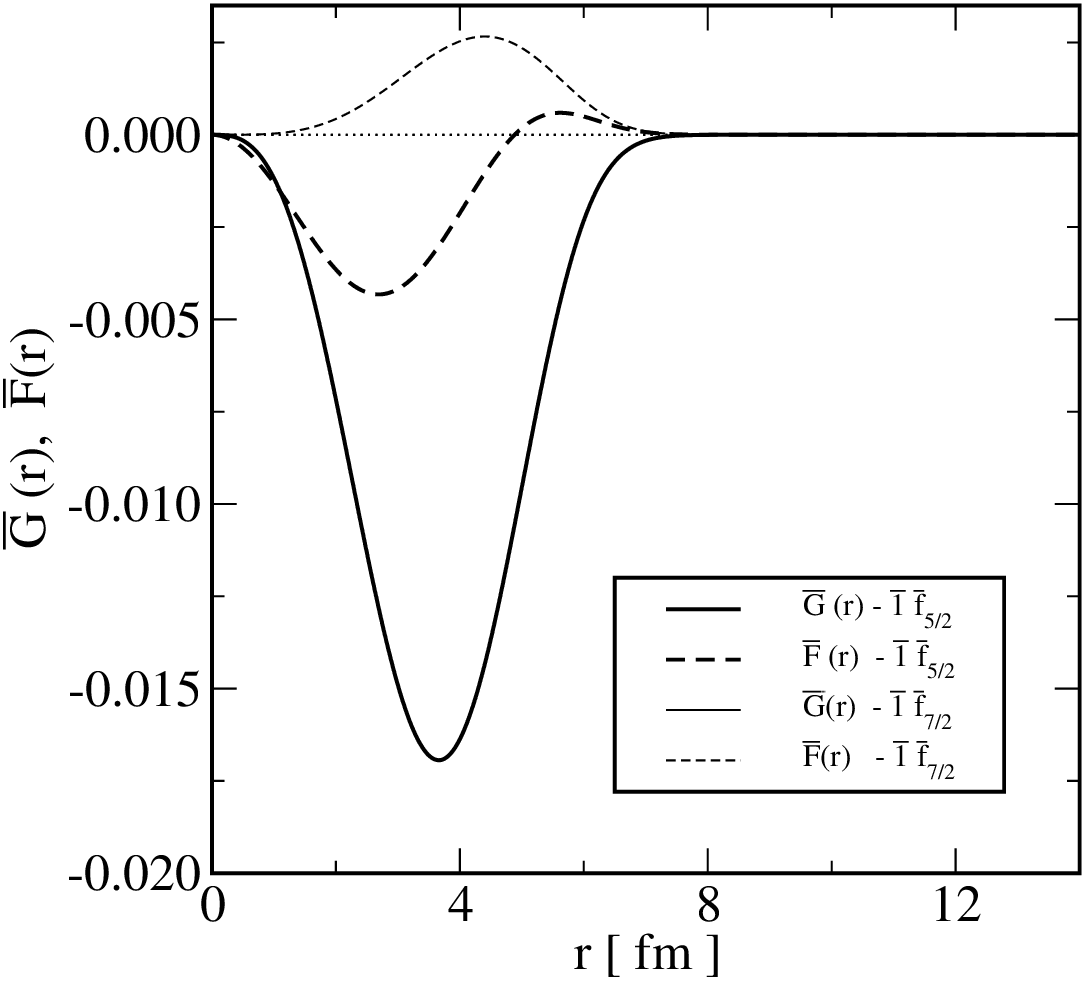}
\end{center}
}\hfill
\parbox[!ht]{8.2cm}{
\begin{center}
\includegraphics[width=8cm]{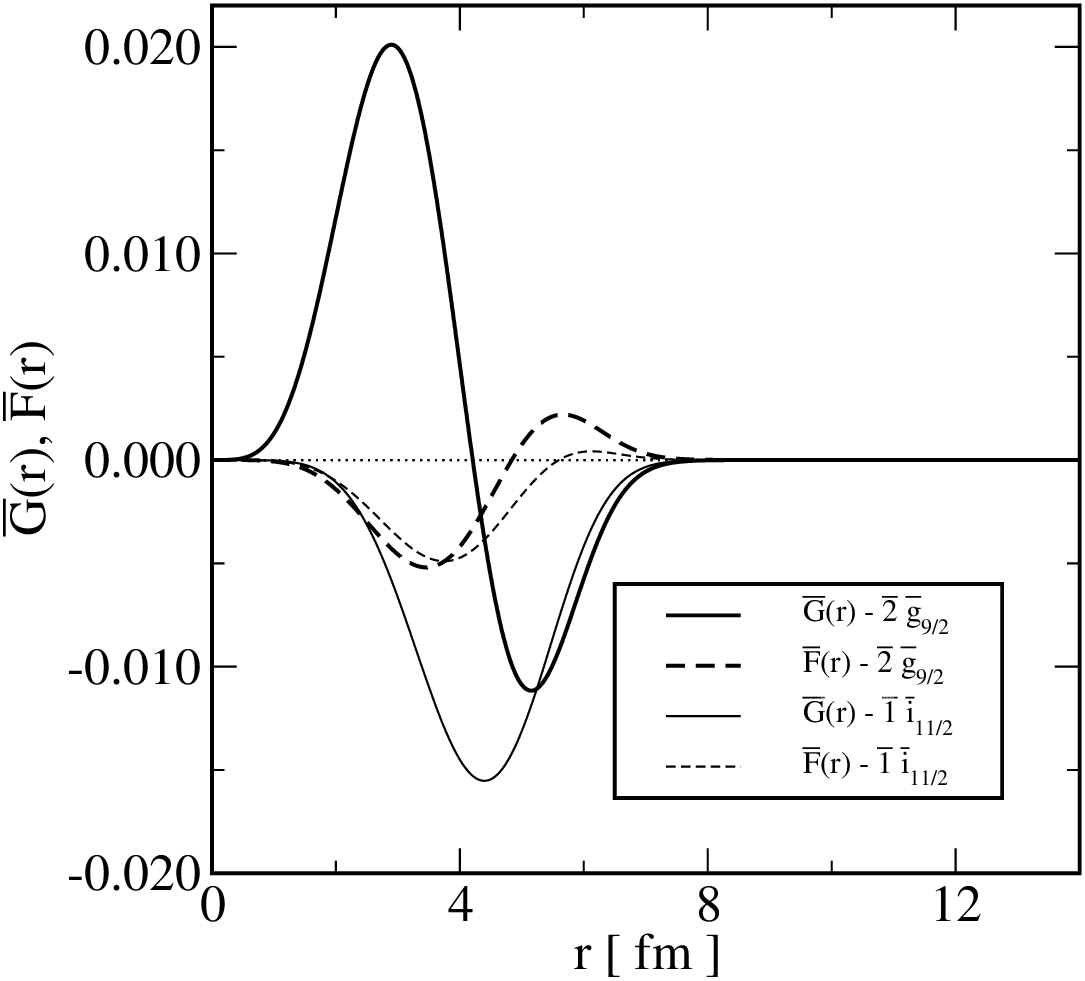}
\end{center} }  } \vspace*{-.2cm}
 \caption{Antineutron radial wave functions $G$ and $F$ of
 the spin pair $\bar 1\bar f_{5/2}-\bar1\bar f_{7/2}$ and the pseudospin pair
 $\bar2\bar g_{7/2}-\bar1\bar i_{11/2}$ for the physical
 parameters in scaled units.\hfill\ } \label{fig:F_G_spin_pspin_an}
\end{figure}


One sees from Fig.~\ref{fig:F_G_spin_pspin_nn} that the spin and pseudospin
pair degeneracy for neutrons is related to the similarity of the $G$ and $F$ wave functions,
respectively. This can be understood by just remarking that the second-order equations for these
functions, having the same quantum numbers for each pair, are basically
the same when the eigenvalues are very similar, which implies, for normalized wave functions,
that the respective solutions should also very similar.

This can also be seen from Fig.~\ref{fig:F_G_spin_pspin_an} for the spin doublet,
which is so much degenerate that the two radial functions $\bar G$ can hardly be
distinguished. Comparing the two figures, one can also see that the amplitude of
the upper radial functions $\bar G$ is bigger for the antineutron case. This is
basically due to the increased kinetic energy of the antineutrons,
which one can roughly estimate by the difference between the depth of
the potential well and the binding energy.

This analysis of the radial functions can only assess the amount of degeneracy of
corresponding doublets and not the perturbative nature of the respective symmetries.
There some possibilies for doing such an analysis using the radial functions,
as was done by Marcos \textit{et al.} \cite{marcos_EPJA26}),
but they are rather indirect. The similarity of the pseudospin partners radial functions $F$ as signature for
pseudospin symmetry was already shown by Ginocchio and Madland in \cite{gino_madland}.

\section{Conclusions}

\label{Sec:conclusions}

In this paper we systematically examined the spin and pseudospin symmetries in nuclei for
realist Woods-Saxon potentials that fit the neutron
single-particle spectrum of $^{208}$Pb. This analysis covers both the nucleon and the antinucleon
spectrum, obtained by charge conjugation of the Dirac equation for mean-field scalar and vector potentials.
By solving the Dirac equation for neutrons and antineutrons for several values of the depth of the vector potential, we were able to perform a systematic analysis of spin and pseudospin symmetries by computing
the energy splittings of the corresponding doublets, as well as computing the spin- and pseudospin-orbit
contributions to those splittings. From that analysis one concludes that spin symmetry, besides being
almost exact for antinucleons, is perturbative and thus can be realized exactly. On the other hand,
pseudospin symmetry, also for nucleons and antinucleons, is found to be not perturbative.
This is probably related with the fact that the potential that drives this symmetry is, in both cases, also the binding potential, and thus one cannot have bound states in the conditions of exact pseudospin symmetry.
However, at this point it is worth remarking that for potentials like the harmonic oscillator is possible to
have exact pseudospin symmetry \cite{prc_65_034307}. A recent study in which a comparison between Woods-Saxon-like
mean-field potentials and harmonic oscillator potentials within a relativistic extension of the Nilsson model
\cite{marcos_EPJA37} gives a hint that, indeed, in the latter case pseudsopin symmetry may be perturbative.

The main difference between the spectrum of single-particle levels of neutrons and antineutrons is the much increased 
depth of the potential (about 10 times as much)
and the kinetic energies. In this sense, the antinucleon bound states are more relativistic than the nucleon
states. However, the main features of the onset of spin and pseudospin symmetry are the same, the differences
being basically quantitative.

\begin{acknowledgments}
We acknowledge financial support from CNPq,
the Capes-FCT project 183/07 and also the projects
PTDC/FIS/64707/2006 and CERN/FP/83505/2008 from FCT.
\end{acknowledgments}

\end{document}